\documentclass{article}
\usepackage[left=2.5cm, right=2.5cm, top=3cm, bottom=3cm]{geometry}
\usepackage{slashed}
\usepackage[utf8]{inputenc}
\usepackage{authblk} 

\usepackage{appendix}
\usepackage{graphicx} 
\usepackage{amsmath}
\usepackage{amssymb}
\usepackage{cite}

\title{Coupling Higher Form Structures of the EFT of Force Free Electrodynamics to Gravity}

\author[a]{Harsh Anand\thanks{harsh.anand@tifr.res.in}}
\affil[a]{Department of Theoretical Physics, Tata Institute of Fundamental Research, Homi Bhabha Rd, Mumbai 400005, India}

\date{May 2026}

\begin{document}

\maketitle

\section{Introduction}

In the study \cite{Gralla:2018oqn} of the EFT of Force Free Electrodynamics, it is observed that the theory admits an effective description in terms of a conserved stress tensor $T^{\mu\nu}$ and a conserved two-form current $J^{\mu\nu}$. These quantities are defined by, for the action $S$,
\begin{equation}\label{def}
    T^{\mu\nu}
\equiv
\frac{2}{\sqrt{-g}}
\frac{\delta S}{\delta g_{\mu\nu}},
\qquad
J^{\mu\nu}
\equiv
\frac{2}{\sqrt{-g}}
\frac{\delta S}{\delta b_{\mu\nu}} .
\end{equation}
Here, the background two-form field $b_{\mu\nu}$ couples to the conserved two-form current associated with magnetic flux conservation and let $a$ be a one form field. Requiring the condition $ db =0$ and the following higher form symmetry: $b \rightarrow b~ +~ d\Lambda ,~~ a \rightarrow a~ +~ \Lambda$ in the effective action reproduces the same long-distance physics and generalized global symmetry structure as microscopic QED within their common regime of validity \footnote{see section \ref{micro}} i.e. their functional path integrals are same. Here, $\Lambda$ can be any arbitrary one form, unlike in QED where $\Lambda$ must be a differential one form ($\Lambda = \partial \phi$ for some scalar $\phi$). The action in QED contains $da = \partial_\mu a_\nu - \partial_\nu a_\mu $; which has the gauge symmetry acting as $a \rightarrow a~+ \partial\phi$, with $\partial\phi = \Lambda$ playing the role of the transformation parameter. 

The simplest action and the leading order term that one can write down for this theory contains simply the term $(b - da)^2$ from which the symmetry mentioned above can be clearly seen. The presence of the two form $b$ helps cancel the arbitrary variation in $a$. \footnote{The reader looking for information on Magnetohydrodynamics and Force Free Electrodynamics may find a brief summary (that is required to understand this paper) in \ref{app}; which has been formed from consulting the sources \cite{Bhattacharyya:2007vjd,
Uchida1997,
Komissarov2002,
GrallaJacobson2014,
GrozdanovHofmanIqbal2017,
FreytsisGralla2016,
Dubovsky2012,
Gaiotto2015,
Gaiotto2018,
HofmanIqbal2018,
Lake2018,
GrozdanovPoovuttikul2017,
GrozdanovLucasPoovuttikul2018,
ArmasJain2018a,
ArmasJain2018b,
GrozdanovPoovuttikul2018,
GloriosoSon2018,
CompereGrallaLupsasca2016,
Kovtun2014,
CrossleyGloriosoLiu2015,
Haehl2016a,
Haehl2016b
} and \cite{Gralla:2018oqn}.}

In this work, we derive a charged spacetime solution from the Einstein-Hilbert action with negative cosmological constant and with the Maxwell field-strength contribution replaced by the gauge-invariant combination
$$
F_{\mu\nu}F^{\mu\nu}
\rightarrow
(b_{\mu\nu}-\partial_\mu a_\nu+\partial_\nu a_\mu)^2.
$$
This model is motivated by the higher-form symmetry structure in the EFT description of Force-Free Electrodynamics. Thus the metric derived from the action will be a solution to the higher form extension of Einstein - Maxwell theory; that is, a space time solution to the leading order EFT of Einstein - Force Free Electrodynamics.

That is, we derive the metric solution from the following Einstein Hilbert action

\[
I
=
-\frac{1}{16\pi G_M}
\int d^{n+1}x \,\sqrt{-g}
\left(
R
-
F^2
+
\frac{n(n-1)}{l^2}
\right)
\]

with
\[
\Lambda
=
-\frac{n(n-1)}{2l^2},
\]
the cosmological constant\footnote{The cosmological constant and the transformation parameter have the same notation $\Lambda$.} associated with the
characteristic length scale \(l\) and 

$$ F = b - da = b_{\mu \nu} - \partial_\mu a_\nu + \partial_\nu a_\mu,$$

which is manifestly different from the field strength $$F = da = \partial_\mu a_\nu - \partial_\nu a_\mu$$

\section{The role of Gravity}

\subsection{The Einstein Hilbert action}

For space time dimension $n+1$ we start our study from the following FFE action:
\begin{equation}\label{action}
I
=
-\frac{1}{16\pi G_M}
\int d^{n+1}x \,\sqrt{-g}
\left(
R
-
F^2
+
\frac{n(n-1)}{l^2}
\right)
\end{equation}

with
\[
\Lambda
=
-\frac{n(n-1)}{2l^2},
\]
the cosmological constant associated with the
characteristic length scale \(l\) and 

$$ F = b - da = b_{\mu \nu} - \partial_\mu a_\nu + \partial_\nu a_\mu,$$

where $b$ is a two form with $db = 0$ and $a$ is a one form. Both of them play the roles of fields present in the action and which combine to give the gauge symmetry \begin{equation}\label{symme}
    b \rightarrow b~ +~ d\Lambda ,~~ a \rightarrow a~ +~ \Lambda.
\end{equation}
We form solutions of the metric from this action in the rest of this paper.

A brief section for the motivation for coupling the theory of Force Free Electrodynamics to gravity may be found in \ref{motiv}. It is hoped that the mentioned motivation may be improvised and broadened in the future. 

\subsubsection{Diffeomorphism invariance of an action containing $\mathcal{F} = b-da$}

Defining
\[
\mathcal{F} \equiv b-da ,
\]
consider the part $S$ of the action $I$ written as
\[
S \sim \int \mathcal{F}\wedge \star \mathcal{F}.
\]
Since \(b\) and \(da\) are both two-forms, \(\mathcal{F}\) is also a two-form, and hence
\[
\mathcal{F}\wedge \star \mathcal{F}
\]
is a top form in four dimensions. \footnote{A top form is a differential form whose degree equals the dimension of spacetime. In \(d\) dimensions, for any \(p\)-form \(\omega\),
\[
\omega \wedge \star \omega
=
\frac{1}{p!}
\omega_{\mu_1 \dots \mu_p}\omega^{\mu_1 \dots \mu_p}
\sqrt{-g}\,d^dx.
\]
Thus \(\omega \wedge \star \omega\) is always a \(d\)-form, i.e. a top form, proportional to the spacetime volume form
\[
\sqrt{-g}\,d^dx.
\]
Since top forms can be integrated directly over spacetime, actions of the form
\[
S \sim \int \omega \wedge \star \omega
\]
are automatically diffeomorphism invariant.} Integrating a top form over spacetime yields a diffeomorphism invariant quantity. Equivalently,
\[
\mathcal{F}\wedge \star \mathcal{F}
=
\frac12
\mathcal{F}_{\mu\nu}\mathcal{F}^{\mu\nu}
\sqrt{-g}\,d^4x,
\]
which is manifestly covariant.

\subsection{The ansatz for the metric in $n+1 = 4$ gravity}

We have the coordinates: $t,r,\theta,\phi$.

The spacetime is static and spherically symmetric, with spatial sections foliated by two-spheres \(S^2\):
\[
\mathcal{M}
\equiv
\mathbb{R}_t \times \mathbb{R}_r \times S^2 .
\]

Accordingly, the metric ansatz is
\[
ds^2
=
- e^{2A(r)} dt^2
+
e^{2B(r)} dr^2
+
r^2 d\Omega_2^2 ,
\]
where
\[
d\Omega_2^2
=
d\theta^2
+
\sin^2\theta\, d\phi^2
\]
is the metric on the unit two-sphere \(S^2\).

\subsection{The ansatz for $b$ and $a$}

\begin{enumerate}
    \item In the paper \cite{Chamblin:1999tk}, the authors derive the charged black hole in an $AdS$ space time and consider just the $a_t$ component of the vector $a_\mu$. Also, this function $a_t$ depends just on the coordinate $r$. This choice is because of the spherical symmetry and staticity in the ansatz of the metric. We shall do the same. That is $$a_\mu = a_t dt = \phi(r) dt.$$ 
    Thus, $$da = \frac{1}{2}(\partial_\mu a_\nu - \partial_\nu a_\mu)~ dx^\mu \wedge dx^\nu= (\partial_\mu \phi(r) - \partial_t a_\mu) ~dx^\mu\wedge dt = (\partial_r \phi(r) )~ dr \wedge dt$$
\item The field $b_{\mu \nu}$ is a two form $b_{\mu \nu}~ dx^\mu \wedge dx^\nu$. Since we require $db = 0$, that is a closed $b$, we may take:

\begin{itemize}
    \item $b$ to be exact: $$b_{\mu \nu} = \partial_\mu c_\nu - \partial_\nu c_\mu$$ for any one form $c_\mu$ or;
    \item 
     a closed but not globally exact two-form; for example consider the Dirac monopole type contribution in $b$:
\begin{equation}\label{ga2}
b = b(r,t)~ dr \wedge dt + q \sin\theta\, d\theta \wedge d\phi .
\end{equation}

This preserves the spherical symmetry of the manifold and, we may verify that
\begin{equation}
db = 0.
\end{equation}
However,
\begin{equation}
\int_{S^2} b
=
q \int_{S^2} \sin\theta\, d\theta \wedge d\phi
=
4\pi q .
\end{equation}
If $ q \neq 0$, this integral is non-vanishing. Therefore, by Stokes' theorem, $b$ cannot be written globally as $b=d\Lambda$, since otherwise
\begin{equation}
\int_{S^2} b
=
\int_{S^2} d\Lambda
=
\int_{\partial S^2} \Lambda
=
0.
\end{equation}
Hence $b$ is a closed but not globally exact two-form.
\end{itemize}
\end{enumerate}

We choose the gauge \eqref{ga2}. This is because a closed and exact two form $b = d\Lambda$ in the action \eqref{action} would not produce any solution different from the RN metric: the action then is just a rewriting of the ordinary Einstein Maxwell action. Hence, we may choose $b = q \sin\theta\, d\theta \wedge d\phi $ and study the Einstein-FFE action and its solutions. For our complete satisfaction however, we study the theory with the gauge \eqref{ga2} because it is the most general two form with $db=0$, respects the spherical symmetry of the metric and is not globally exact. The static nature  of the metric would further constrain the dependence of $b(r,t)$ on $t$ to vanish but we derive this requirement from the equations of motion in the following sections.

\subsubsection{The solution of this gauge theory is the dyonic Reissner Nordstr\"{o}m metric}\label{impsect}

If the background two-form $b$ is exact, namely
\[
b=dc
\]
for some globally defined one-form $c$, then the field redefinition
\[
a' = a-c
\]
transforms the action
\[
S=\int (b-da)^2
\]
into
\[
S=\int (da')^2.
\]
Thus, the theory is simply the same as ordinary Maxwell theory, and the two actions have the same solutions.

However, if $b$ is closed but not globally exact, then no globally defined one-form $c$ exists such that $b=dc$. In this case the above field redefinition cannot be performed globally. Note that it can still be performed locally. In any sufficiently small patch we may write
\[
b=d\lambda ,
\]
so that
\[
b-da=d(\lambda-a).
\]
Locally, the action therefore takes the same form as the Maxwell action after the redefinition $A=\lambda-a$. And thus, the local equations of motion are identical to those of Maxwell theory.

But there is a difference that appears only at the global level. The combination
\[
F=b-da
\]
can carry a nonzero flux through the two-sphere coming from $b$,
\[
\int_{S^2}F=\int_{S^2}b,
\]
and for a globally defined gauge potential we always has
\[
\int_{S^2}da=0
\]
by Stokes' theorem. Since the value of this flux is a global property of the field configuration, a solution with nonzero flux cannot be continuously deformed into one with zero flux. Therefore a field configuration solving the $(b-da)^2$ theory generally cannot be represented as a solution of the $(da)^2$ theory. Thus, the two theories obey the same local equations but belong to different global sectors, leading to different sets of classical solutions. 

We shall see in the following sections that this is true. However, we note that the solution we obtain is the same as a dyonic Reissner Nordstr\"{o}m metric which is the solution of the Einstein-Maxwell theory with $F= da$ and $a$ having a Dirac monopole; that is, $F= q \sin \theta ~d\theta \wedge d\phi$.

\subsection{The stress tensor}\label{str}

We have $$F^2 = F_{\rho \sigma}F^{\rho \sigma}$$ $$F^{\rho \sigma} = g^{\rho \alpha}g^{\sigma \beta} F_{\alpha \beta}$$

Thus, $$\delta(\sqrt{-g} F^2) = \sqrt{-g}\left( 2F_{\mu \rho}F_{\nu}^{\rho} - \frac{1}{2} g_{\mu \nu} F^2 \right) $$

so that the stress tensor is $$T_{\mu \nu}= 2 \left( F_{\mu \rho}F_{\nu}^{\rho} - \frac{1}{4} g_{\mu \nu} F^2 \right)$$

\subsubsection{The conservation of the stress tensor}

The conservation of the Maxwell stress tensor,
\begin{equation}
\nabla_\mu T^{\mu\nu}=0,
\end{equation}
requires only the source-free Maxwell equations, and the Bianchi identity: 
\begin{equation}
\nabla_\mu F^{\mu\nu}=0,
\end{equation}

\begin{equation}
dF=0
\qquad\left(\text{equivalently } \nabla_{[\mu}F_{\nu\rho]}=0\right).
\end{equation}
Thus, the stronger condition that the $F$ be globally exact,
\begin{equation}
F=dc,
\end{equation}
is not required. \footnote{For example, in the presence of a Dirac monopole, $F$ is closed but not globally exact, i.e.
\begin{equation}
dF=0,
\qquad
F\neq dc,
\end{equation}
while the stress tensor remains covariantly conserved.}

In the case of study in this paper, we have considered $F = b-da$ with $dF = 0$ but $F \neq dc$ for any $c$ and thus, we the Maxwell equation enforces automatically the conservation of the stress tensor.

\subsection{The final Einstein equation}\label{gmu}

By standard techniques, we build the Einstein equation:

\begin{equation}\label{Ein}
    R_{\mu \nu} - \frac{1}{2}R g_{\mu \nu} + \Lambda g_{\mu \nu} = T_{\mu \nu} = 2\left(F_{\mu \rho}F_{\nu}^{\rho} - \frac{1}{4} g_{\mu \nu} F^2 \right)
\end{equation}

We have an ansatz for the metric, for the two form $b$ and the one form $a$. Thus we may use \eqref{Ein} to solve for these fields. But the $b,a$ appearing in $F_{\mu \nu}$ are constrained by the analogue of the Maxwell equations derived below.

\subsection{Equation of motion from varying $a_\mu$}

We have a part of the action 

\begin{equation}
    S = \int d^4 x \sqrt{-g} \left( b_{\mu \nu} - \partial_\mu a_\nu + \partial_\nu a_\mu \right)\left( b^{\mu \nu} - \partial^\mu a^\nu + \partial^\nu a^\mu \right)
\end{equation}

The equation of motion is just the analogue of the Maxwell's equation $\nabla_\mu F^{\mu\nu} = 0 $ where $F$ is $b-da$. It can be written as:

\begin{equation}\label{Max}
    \frac{1}{\sqrt{-g}} \partial_\mu \left(\sqrt{-g} \left( b^{\mu \nu} - \partial^\mu a^\nu + \partial^\nu a^\mu \right)\right) = 0 
\end{equation}

\subsubsection{Solving the above}

We have,

$$ds^2
=
- e^{2A(r)} dt^2
+
e^{2B(r)} dr^2
+
r^2 d\Omega_2^2$$

For the ansatz
\begin{equation}
a_\mu dx^\mu = \phi(r)\,dt,
\qquad
b = b_{\mu\nu}~dx^\mu\wedge dx^\nu = b(r,t)~ dr \wedge dt + Q_m \sin\theta\, d\theta \wedge d\phi
\end{equation}
the field strength, is
\begin{equation}
F_{\mu\nu}~dx^\mu\wedge dx^\nu
=
b-da
=
q\sin\theta\, d\theta\wedge d\phi +
(b(r,t) -\phi'(r))\,dr\wedge dt.
\end{equation}

The non-vanishing components of $F_{\mu\nu}$ are therefore
\begin{equation}
F_{tr}=\phi'(r)- b(r,t),
\qquad
F_{\theta\phi}=q\sin\theta,
\end{equation}
together with those related by antisymmetry.

Raising the indices using the metric,
\begin{equation}
F^{\mu\nu}
=
\begin{pmatrix}
0 &
e^{-2\left(A(r)+B(r)\right)}
\left(b(r,t)-\phi'(r)\right)
&
0 &
0
\\[6pt]
e^{-2\left(A(r)+B(r)\right)}
\left(\phi'(r)-b(r,t)\right)
&
0 &
0 &
0
\\[6pt]
0 &
0 &
0 &
\dfrac{q\csc\theta}{r^4}
\\[8pt]
0 &
0 &
-\dfrac{q\csc\theta}{r^4}
&
0
\end{pmatrix}.
\end{equation}

\begin{equation}\label{maxsol}
\nabla_\mu F^{\mu\nu}
=
\begin{pmatrix}
\dfrac{e^{-\left(A(r)+B(r)\right)}}{r^2}
\dfrac{d}{dr}
\!\left[
r^2 e^{-\left(A(r)+B(r)\right)}
\left(\phi'(r)-b(r,t)\right)
\right]
\\[12pt]
e^{-2\left(A(r)+B(r)\right)}
\,\partial_t b(r,t)
\\[8pt]
0
\\[8pt]
0
\end{pmatrix}.
\end{equation}

We anticipated the second line of the above vector to be zero because of the staticity in the anstaz of the metric.
\newline
\newline
\newline
\newline

\textit{In the next page, we give the matrix expressions for the stress tensor and the Einstein tensor.}

\clearpage
\newgeometry{left=1cm, right=1cm, top=1cm} 

\begin{center}

\vspace*{\fill}
\subsection{The stress tensor}
\begin{equation}
T^\mu{}_\nu
=
\begin{pmatrix}
-\;e^{-2\left(A(r)+B(r)\right)}
\left(b(r,t)-\phi'(r)\right)^2
-\dfrac{q^2}{r^4}
&
0
&
0
&
0
\\[8pt]
0
&
-\mathcal{E}
&
0
&
0
\\[8pt]
0
&
0
&
\mathcal{E}
&
0
\\[8pt]
0
&
0
&
0
&
\mathcal{E}
\end{pmatrix}.
\end{equation}

That is,
\begin{equation}
T^\mu{}_\nu
=
\operatorname{diag}
\!\left(
-\mathcal{E},
-\mathcal{E},
\mathcal{E},
\mathcal{E}
\right),
\qquad
\mathcal{E}
=
e^{-2\left(A(r)+B(r)\right)}
\left(b(r,t)-\phi'(r)\right)^2
+\frac{q^2}{r^4}.
\end{equation}
\vspace*{\fill}

\subsection{The Einstein tensor}

The Einstein tensor $G_{\mu\nu}$ is

\[
\begin{pmatrix}
\dfrac{
e^{2A(r)-2B(r)}
\left(
2rB'(r)+e^{2B(r)}-1
\right)
}{r^2}
& 0 & 0 & 0 \\[10pt]

0 &
\dfrac{
2rA'(r)-e^{2B(r)}+1
}{r^2}
& 0 & 0 \\[10pt]

0 & 0 &
r e^{-2B(r)}
\left(
rA''(r)
+
A'(r)\left(1-rB'(r)\right)
+
rA'(r)^2
-
B'(r)
\right)
& 0 \\[10pt]

0 & 0 & 0 &
\sin^2\theta G_{33}
\end{pmatrix}
\]

\vspace*{\fill}

\end{center}

\clearpage 

\restoregeometry 

\subsection{Einstein equation: Combining \ref{str} and \ref{gmu} }

Einstein's equation is \eqref{Ein}. Contracting both sides by the inverse metric, we get for the Einstein tensor $G_{\mu \nu}$.
\begin{equation}\label{phi}
    G^{\mu}{}_{\nu}
+
\Lambda \,\delta^{\mu}{}_{\nu}
=
T^{\mu}{}_{\nu}
\end{equation}

Subtracting the \(tt\) and \(rr\) components gives
\begin{equation}\label{subt}
    G^{t}{}_{t}
-
G^{r}{}_{r}
=
T^{t}{}_{t}
-
T^{r}{}_{r}.
\end{equation}

Now from the matrix obtained for $T_{\mu \nu}$, we find that $$g^{tt}T_{tt} = g^{rr}T_{rr}$$ so that $$T^t_t = T^r_r$$

and thus, the right-hand side of \eqref{subt} vanishes:
\[
G^{t}{}_{t}
-
G^{r}{}_{r}
=
0.
\]

Using the metric ansatz,
\[
G^{t}{}_{t}
-
G^{r}{}_{r}
=
-\frac{2e^{-2B(r)}}{r}
\left(
A'(r)+B'(r)
\right),
\]
and therefore
\[
-\frac{2e^{-2B(r)}}{r}
\left(
A'(r)+B'(r)
\right)
=
0.
\]

Since
\[
e^{-2B(r)} \neq 0,
\]
it follows that
\[
A'(r)+B'(r)=0.
\]

Hence,
\[
A(r) = -B(r) + \text{constant}.
\]

Without loss of generality, we may set the constant to zero or absorb it in the definition of $A(r)$.

Setting $$e^{2A(r)} = f(r),$$

the metric ansatz becomes 

$$ds^2= - f(r) dt^2 + \frac{dr^2}{f(r)} + r^2 d\Omega_2^2$$

\subsection{Solving for $f(r)$}

We notice because
$$T^t_t = T^r_r,$$  we again get the familiar equation $$A(r) + B(r) = 0.$$

This when plugged into \eqref{maxsol},
which was 
\begin{equation}\label{maxsol}
\nabla_\mu F^{\mu\nu}
=
\begin{pmatrix}
\dfrac{e^{-\left(A(r)+B(r)\right)}}{r^2}
\dfrac{d}{dr}
\!\left[
r^2 e^{-\left(A(r)+B(r)\right)}
\left(\phi'(r)-b(r,t)\right)
\right]
\\[12pt]
e^{-2\left(A(r)+B(r)\right)}
\,\partial_t b(r,t)
\\[8pt]
0
\\[8pt]
0
\end{pmatrix}.
\end{equation}

we get \begin{equation}\label{21}
    \phi'(r) - b(r,t) = \frac{Q}{r^2}
\end{equation}

where $Q$ is a constant, independent of $r,t$. $Q$ is independent of $t$ also because of the second line of \eqref{maxsol}.
\newline

Now we must solve for $f(r)$.

We have $$G^t_t + \Lambda = T^t_t = -\left(b(r)- \phi'(r)\right)^2 - \frac{q^2}{r^4}$$

But $$G^t_t= \frac{f'}{r}+ \frac{f-1}{r^2}$$

Thus the differential  equation we have is $$\frac{f'}{r}+ \frac{f-1}{r^2} + \Lambda= -\left(\frac{Q}{r^2}\right)^2 - \frac{q^2}{r^4} $$

It's solution is 

\[
f(r)
=
1+\frac{Q^2+q^2}{r^2}
-\frac{\Lambda r^2}{3}
-\frac{M}{r}.
\]

\section{The Reissner Nordstr\"{o}m metric and the new metric}

\begin{enumerate}
    \item Reissner Nordstr\"{o}m :
    
    $$ds^2= - f_{RN}(r) dt^2 + \frac{dr^2}{f_{RN}(r)} + r^2 d\Omega_2^2$$
 Here, $$f_{RN}(r)
=
1+\frac{Q^2}{r^2}
-\frac{\Lambda r^2}{3}
-\frac{M}{r}$$

Note that the action contained $F^2 = (da)^2$ and the gauge choice used here was $a = a_t dt = \phi(r)dt$. After solving for $\phi(r)$, we get $\phi(r) = -\frac{Q}{r}$, where $Q$ is a constant of integration. 

    \item The metric from Einstein- FFE theory:
    
    $$ds^2= - f(r) dt^2 + \frac{dr^2}{f(r)} + r^2 d\Omega_2^2$$

    Here, $$f(r)=1+\frac{Q^2+q^2}{r^2}
-\frac{\Lambda r^2}{3}
-\frac{M}{r}.$$

Note here that $$ b - da = F = 
q\sin\theta\, d\theta\wedge d\phi +
(b(r,t) -\phi'(r))\,dr\wedge dt = 
q\sin\theta\, d\theta\wedge d\phi -\frac{Q}{r^2}\,dr\wedge dt,$$ from \eqref{21}.

This is thus the dyonic Reissner Nordstr\"{o}m solution. 

\end{enumerate}



\section{Conclusion}

We know that the charged Reissner Nordstr\"{o}m black hole metric is obtained from the Einstein Hilbert gravitational action. This action has the kinetic term $F^2 = (da)^2$. 

However, in order to describe a gravitational theory coupled to the leading order structure of the effective field theory of Force Free Electrodynamics, we propose this change in the action: $F^2= (b-da)^2$. This ensures that the new action has the higher form symmetry \eqref{symme}. The requirement $db = 0$ and the subsequent choice $b = dc$ (that is, exact $b$), for some $c$ makes this theory trivially equivalent to Einstein-Maxwell theory. 
In this work, we choose a closed $b$ but which is not globallly exact, that is $b$ cannot be written as $d\Lambda$ for a one form $\Lambda$, and note that this time the theory is not trivially equivalent to Einstein-Maxwell theory.
As explained in \ref{impsect}, such a $b$ would produce a non zero flux through the two-sphere, carried by $F = b-da$. This is different from a theory with $F=da$ in which $F$ carries zero flux through the two sphere. Here, by flux, we mean the quantity:

$$\int_{S^2} F.$$

However it is similar to an Einstein-Maxwell theory that has a Dirac Monopole. In this theory, $$F = da = q \sin \theta~ d \theta \wedge d\phi$$ and thus closely resembles the theory we considered.

Thus we conclude that the Einstein-FFE theory we considered, has the same solution as the Einstein-Maxwell theory. We have considered the case with a Dirac monopole type term in the $b$ and checked that the theory gives the dyonic  Reissner Nordstr\"{o}m solution, which is the same solution that comes from the Einstein-Maxwell theory when the $a$ in $F= da$ has a Dirac Monopole.

\section*{Acknowledgments}

We are grateful to Shiraz Minwalla for directing me towards the study of Force Free Electrodynamics and Magneto Hydrodynamics, and the possibility of the gravity dual of this theory; and for insightful guidance at every step during the writing of this project. We would also like to thank the department of theoretical physics, our colleagues and especially Pratik Rath, at TIFR for invaluable discussions and encouragement.







\appendix

\section{Brief review of FFE}\label{app}

\subsection{The new action}

Effective FFE  has as it's starting point the conservation of two tensors $T,J$ and the presence of a higher form symmetry in its action (see \eqref{symmetry}). The tensors are defined as:

\begin{equation}\label{def}
    T^{\mu\nu}
\equiv
\frac{2}{\sqrt{-g}}
\frac{\delta S}{\delta g_{\mu\nu}},
\qquad
J^{\mu\nu}
\equiv
\frac{2}{\sqrt{-g}}
\frac{\delta S}{\delta b_{\mu\nu}} .
\end{equation}

The two conditions, that a theory has two conserved tensors and that the theory has the higher form symmetry are equivalent to each other, as we now prove (the proof has been obtained from \cite{Gralla:2018oqn}).

\subsubsection{The role of the higher form symmetry}

We shall prove that an action with the symmetry:
\begin{equation}\label{symmetry}
    b \rightarrow b + d\Lambda,~~~ a \rightarrow a +\Lambda
\end{equation}
has the above mentioned two conserved currents as its equations of motion in the subsection \eqref{eqofa} below. That is, that equations \eqref{topro} are a consequence of having the higher form symmetry, where these equations are:

\begin{equation}\label{topro}
\nabla_\mu T^{\mu\nu}
=
\frac{1}{2}
(db)^{\nu}{}_{\rho\sigma}
J^{\rho\sigma} = 0,
\qquad
\nabla_\mu J^{\mu\nu}
=
0 .
\end{equation}

In \cite{Gralla:2018oqn}, it is stated that an action with this higher form symmetry \eqref{symmetry}, is equivalent to having two conserved currents as a result of the equations of motion formed from the action.

One direction of the equivalence, that is that the presence of the symmetry \eqref{symmetry} leads to the action having the two conserved currents as the equations of motion, is proved in \eqref{eqofa}.

The full equivalence actually also demands the converse to be proven, that having the two conserved currents in a theory with an action as its equation of motions leads to the theory having the symmetry \eqref{symmetry} and has been explained in the section "B. The field equations are conservation laws" in \cite{Gralla:2018oqn}.

\subsubsection{The equation of motion of $a_\mu$ in an action possessing the symmetry \eqref{symmetry}}\label{eqofa}

Let the action contain the metric, the two form field $b_{\mu \nu}$, the one form field $a_\mu$ and some fields $\Phi$.

Using the Lie derivatives
\begin{align}
\delta_\xi g_{\mu\nu} &= \mathcal{L}_\xi g_{\mu\nu}, \nonumber\\
\delta_\xi b_{\mu\nu} &= \mathcal{L}_\xi b_{\mu\nu}, \nonumber\\
\delta_\xi a_\mu &= \mathcal{L}_\xi a_\mu, \nonumber\\
\delta_\xi \Phi^I &= \mathcal{L}_\xi \Phi^I,
\end{align}
the diffeomorphism variation
\[
\delta_\xi S
=
\int d^4x
\left(
\frac{\delta S}{\delta g_{\mu\nu}}\delta_\xi g_{\mu\nu}
+
\frac{\delta S}{\delta b_{\mu\nu}}\delta_\xi b_{\mu\nu}
+
\frac{\delta S}{\delta a_\mu}\delta_\xi a_\mu
+
\frac{\delta S}{\delta\Phi^I}\delta_\xi\Phi^I
\right)
\]
can be rewritten as
\[
\delta_\xi S
=
\int d^4x\,\sqrt{-g}
\left[
-\nabla_\mu T^\mu{}_{\sigma}
+\frac12 J^{\mu\nu}(db)_{\sigma\mu\nu}
-\nabla_\mu J^{\mu\nu}\,b_{\sigma\nu}
\right]\xi^\sigma
+\frac{\delta S}{\delta a_\mu}\delta_\xi a_\mu
+\frac{\delta S}{\delta\Phi^I}\delta_\xi\Phi^I.
\]

where, the variation of the metric gives the first term, the variation of the two form $b$ in the action gives the second and the third term (where $J^{\mu\nu}
=
\frac{2}{\sqrt{-g}}
\frac{\delta S}{\delta b_{\mu\nu}}$ and $T^{\mu \nu}$ is the stress tensor).

Now, if the action has the symmetry 
\begin{equation}
    b \rightarrow b + d\Lambda, a \rightarrow a +\Lambda
\end{equation}

That is, action is invariant under the one-form gauge transformation
\begin{equation}
b_{\mu\nu}\rightarrow b_{\mu\nu}+2\partial_{[\mu}\Lambda_{\nu]},
\qquad
a_\mu\rightarrow a_\mu+\Lambda_\mu,
\end{equation}
under which
\begin{equation}
\delta_\Lambda S
=
\int d^4x
\left(
2\frac{\delta S}{\delta b_{\mu\nu}}
\partial_{[\mu}\Lambda_{\nu]}
+
\frac{\delta S}{\delta a_\mu}\Lambda_\mu
\right).
\end{equation}

Integrating the first term by parts, and neglecting boundary terms, gives
\begin{align}
\delta_\Lambda S
&=
\int d^4x
\left(
-2\partial_\mu
\left(
\frac{\delta S}{\delta b_{\mu\nu}}
\right)
\Lambda_\nu
+
\frac{\delta S}{\delta a_\nu}\Lambda_\nu
\right)
\nonumber\\
&=
\int d^4x
\left[
-2\partial_\mu
\left(
\frac{\delta S}{\delta b_{\mu\nu}}
\right)
+
\frac{\delta S}{\delta a_\nu}
\right]
\Lambda_\nu .
\end{align}

Since $\Lambda_\mu$ is arbitrary, gauge invariance implies the Noether identity
\begin{equation}
\frac{\delta S}{\delta a_\nu}
=
2\partial_\mu
\left(
\frac{\delta S}{\delta b_{\mu\nu}}
\right).
\end{equation}

We have, the two-form current
\begin{equation}
J^{\mu\nu}
=
\frac{2}{\sqrt{-g}}
\frac{\delta S}{\delta b_{\mu\nu}},
\end{equation}
the identity becomes
\begin{equation}
\frac{1}{\sqrt{-g}}
\frac{\delta S}{\delta a_\nu}
=
\nabla_\mu J^{\mu\nu}.
\end{equation}

Therefore, the equation of motion for $a_\mu$,
\begin{equation}
\frac{\delta S}{\delta a_\mu}=0,
\end{equation}
is equivalent to the conservation law
\begin{equation}
\nabla_\mu J^{\mu\nu}=0.
\end{equation}

Thus the equations of motion are $$-\nabla_\mu T^\mu{}_{\sigma}
+\frac12 J^{\mu\nu}(db)_{\sigma\mu\nu} = 0$$
and $$\nabla_\mu J^{\mu\nu} = 0$$

If the stress tensor is conserved, $$\nabla_\mu T^\mu{}_{\sigma} = \frac12 J^{\mu\nu}(db)_{\sigma\mu\nu} = 0$$

Thus, we have shown that the consequence of a theory possessing the symmetry \eqref{symmetry} is that the theory has two conserved tensor currents.

We now consider in \eqref{micro} an example of an action that has the higher form symmetry and thus has the two conserved tensor currents. This theory is the microscopic QED.

\subsection{Microscopic QED}\label{micro}

We refer mainly to \cite{Gralla:2018oqn, Glorioso:2018kcp,Grozdanov:2016tdf, Iqbal:2025dsh} for this section. In the work \cite{Gralla:2018oqn}, the authors start with the partition function of the microscopic QED:

\begin{equation}
    Z[g,b] = \int [d\psi dA] \exp(iS_{QED}[\psi,A;g,b])
\end{equation}

where, $S_{\mathrm{QED}}$ is the microscopic QED action:

\begin{equation}\label{2}
    S_{\mathrm{QED}}
=
\int d^4x \,\sqrt{-g}\,
\left[
-\frac{1}{4}(dA)^2
+
\bar{\psi}\,(\slashed{D}+m)\,\psi
+
\frac{1}{4}\,
b_{\mu\nu}\,\epsilon^{\mu\nu\rho\sigma}(dA)_{\rho\sigma}
\right]
\end{equation}

In \eqref{2}, these are the symmetries present:

\begin{equation}\label{symm}
    \begin{split}
    Z[\phi^* g,\phi^* b] &= Z[g,b]\\
Z[g,b+d\Lambda] &= Z[g,b]
\end{split}
\end{equation}

Each symmetry is associated with a conservation law.
Varying with respect to an infinitesimal 1-form shift
\[
b \to b + d\Lambda ,
\]
 we find, by integration by parts of the last term of the action,
\[
\nabla_\mu \langle J^{\mu\nu} \rangle = 0 ,
\]

where, $$J^{\mu \nu} = \epsilon^{\mu\nu\rho\sigma}(dA)_{\rho\sigma}.$$

After varying the partition function with respect to an infinitesimal diffeomorphism and then using this result, it follows from the two symmetries that
\[
\nabla_\mu \langle T^{\mu\nu} \rangle
=
\frac{1}{2}(db)^{\nu}{}_{\rho\sigma}
\langle J^{\rho\sigma} \rangle .
\]

The right-hand side reflects non-conservation in the
presence of an external electric current \(db\). Thus, in
the absence of external electric charge, the symmetries imply the conservation of the correlators ,
\[
\nabla_\mu \langle T^{\mu\nu} \rangle = 0,
\qquad
\nabla_\mu \langle J^{\mu\nu} \rangle = 0 .
\]

This form of presenting the QED minimises the importance of the degrees of freedom $(\psi, A)$ that are present in the measure of the partition function.

In more common presentations of QED, the $J_{\mu \nu}$ is defined as the Hodge star of the field strength $F_{\mu \nu}$ and is conserved following the Bianchi identity acting on the Maxwell equations. In comparison, here, in the action \eqref{2} the $J_{\mu \nu}$ is conserved due to the presence of the antisymmetric two form $b_{\mu \nu}$ present in the action coupled to the Hodge star of the field strength $dA$; with the symmetry described in \eqref{symm}. 

Note that the Bianchi identity acting on the Maxwell equations is just the statement $dF = d(dA) = 0$ which in comparison is indeed used here in the form of $dF = d(b - da) = 0, ~~~ \rm{as} ~~ db = 0$.

Also, in comparison, in this action \eqref{2}, the stress tensor $T_{\mu \nu}$ is conserved due to diffeomorphism invariance as well as the vanishing of the external source $b_{\mu \nu}$.



\subsection{Force Free Electrodynamics}\label{motiv}

The calculations in \cite{Gralla:2018oqn} conclude that "conventional force-free electrodynamics is the unique scale-free theory describing the infrared dynamics of cold string fluids." Thus, the theory of Force Free Electrodynamics coupled to gravity is in itself an interesting field of study; this is because we have learnt that FFE is the infrared study of cold strings, which are magnetic field lines behaving almost like physical objects, having twists and loops. Very much like hydrodynamics is a coarse grained theory of elementary particles, FFE is the coarse graining of micro quantum electro dynamics.

What information could be contained in the calculation of such a theory coupled to gravity? We shall explore it in this paper.

\bibliographystyle{unsrt}
\bibliography{Ref}

\end{document}